\documentclass[%
 reprint,
superscriptaddress,
amsmath,amssymb,
 aip,
]{revtex4-1}
\usepackage{graphicx}
\usepackage{dcolumn}
\usepackage{bm}
\usepackage{physics} 
\usepackage{xcolor}
\usepackage{textcomp}
\usepackage[utf8]{inputenc}
\usepackage[T1]{fontenc}
\begin{document}

\preprint{AIP/123-QED}

\title{Optically Tailored  Trapping Geometries for Ultracold Atoms on a Type-II Superconducting Chip}
\author{Francesca Tosto}
\affiliation{%
 Centre for Quantum Technologies, National University of Singapore, 3 Science Drive 2, Singapore 117543, Singapore
 }%
\author{Phyo Baw Swe}
\affiliation{%
 Division of Physics and Applied Physics, Nanyang Technological University, 21 Nanyang Link, Singapore 637371, Singapore
 }%
\author{Nghia Tin Nguyen}
\affiliation{%
 Division of Physics and Applied Physics, Nanyang Technological University, 21 Nanyang Link, Singapore 637371, Singapore
 }%
\author{Christoph Hufnagel}
\affiliation{%
 Centre for Quantum Technologies, National University of Singapore, 3 Science Drive 2, Singapore 117543, Singapore
 }%
\author{Mar\'ia Mart\'inez Valado}
\affiliation{%
 Centre for Quantum Technologies, National University of Singapore, 3 Science Drive 2, Singapore 117543, Singapore
 }%
\author{Leonid Prigozhin}
\affiliation{J. Blaustein Institutes for Desert Research, Ben-Gurion University of the Negev, Sede Boqer Campus 84990, Israel
 }%
 \author{Vladimir Sokolovsky}
\affiliation{Physics Department, Ben-Gurion University of the Negev, Beer Sheva 84105, Israel
 }%
\author{Rainer Dumke} 
\affiliation{%
 Centre for Quantum Technologies, National University of Singapore, 3 Science Drive 2, Singapore 117543, Singapore
 }%
 \affiliation{%
 Division of Physics and Applied Physics, Nanyang Technological University, 21 Nanyang Link, Singapore 637371, Singapore
 }%

\date{\today}

\begin{abstract}
Superconducting atom chips have very significant advantages in realizing trapping structures for ultracold atoms compared to conventional atom chips. We extend these advantages further by developing the ability to dynamically tailor the superconducting trap architecture.
Heating the chosen parts of a superconducting film by transferring optical images onto its surface we are able to modify the current density distribution and create desired trapping potentials. This method enables us to change the shape and structure of magnetic traps, enabling versatile applications in atomtronics.
\end{abstract}

\maketitle

In the past years, superconducting (SC) atom chips have drawn a lot of attention for trapping and manipulating ultracold atoms \cite{2040-8986-18-9-093001,10.1103/PhysRevA.76.033618,10.1103/PhysRevLett.101.183006,09500340.2016.1178820,book,10.1038/ncomms3380,10.1063/1.4852017,10.1103/PhysRevLett.114.113003,2ndbook}. Due to the suppression of magnetic field noise close to superconducting surfaces, SC atom chips are excellent candidates for quantum information applications \cite{PhysRevLett.97.070401,PhysRevA.76.033618,10.1140/epjd/e2009-00001-5,10.1209/0295-5075/87/13002,0953-4075-43-9-095002,PhysRevA.72.042901,PhysRevA.75.064901,Dikovsky2009}. Moreover they will enable the realization of hybrid quantum systems composed of ultracold atoms and superconducting circuits, thereby merging the advantages of these two technologies \cite{sch,10.1103/PhysRevA.90.040502,2015_n,PhysRevA.93.042329,PhysRevA.94.062301,2058-9565-2-3-035005,1367-2630-20-2-023031,fir,fif,fou,fou1,sec,thi}.  

Owing to the characteristic properties of superconductors, magnetic traps for the confinement of neutral atoms on a SC atom chip can be created by either transport currents \cite{PhysRevLett.97.200405,0295-5075-81-5-56004,10.21468/SciPostPhys.4.6.036}, persistent currents \cite{PhysRevLett.98.260407,10.1103/PhysRevA.79.053641,10.1007/s00340-014-5768-3}, or trapped magnetic fields induced by pinned vortices \cite{10.1103/PhysRevLett.103.253002,PhysRevA.81.053624,1367-2630-12-4-043016}. For transport and persistent currents the trap type is determined by the shape of the current carrying wire and therefore cannot be changed during experiments.
Magnetic traps generated by vortices inside type-II superconductors allow a more flexible approach. Depending on the history of the applied magnetic fields and transport currents, various traps with or without external bias fields can be created by the same superconducting structures. In addition, such traps are easier to manipulate using external magnetic fields. These  traps are, however, limited to simple geometries  and  do not allow one to change the distribution of vortices locally \cite{0953-2048-27-12-124004,Sol2,Sol1}. 

In this paper we describe in detail how to configure the vortex distribution in a square SC film in order to create the desired potential without the need of changing the chip or the applied field. We make use of a high-power laser and a DMD (digital micromirror device) to destroy the superconductivity in  selected regions of the film and influence the shape and structure of a trap.
We have been able to realize various trapping potentials and, in particular, to split a single trap or to transform it into a crescent or a ring-like trap. Since the atomic cloud evolves  with the trapping potential, cold atoms can be used as a sensitive probe to examine the real-time magnetic field and vortex distribution. We  experimentally verified the possibility of controlling  trap geometry by local laser heating of a SC film. We have also carried out simulations of the film heating, the corresponding redistribution of sheet current density, and the induced trapping potentials; the simulation results are in agreement with the experiments. Such simulations help to better understand the process and can be used to design a trap with the needed properties. \\
\begin{figure}[t!]
  \includegraphics[scale=0.3]{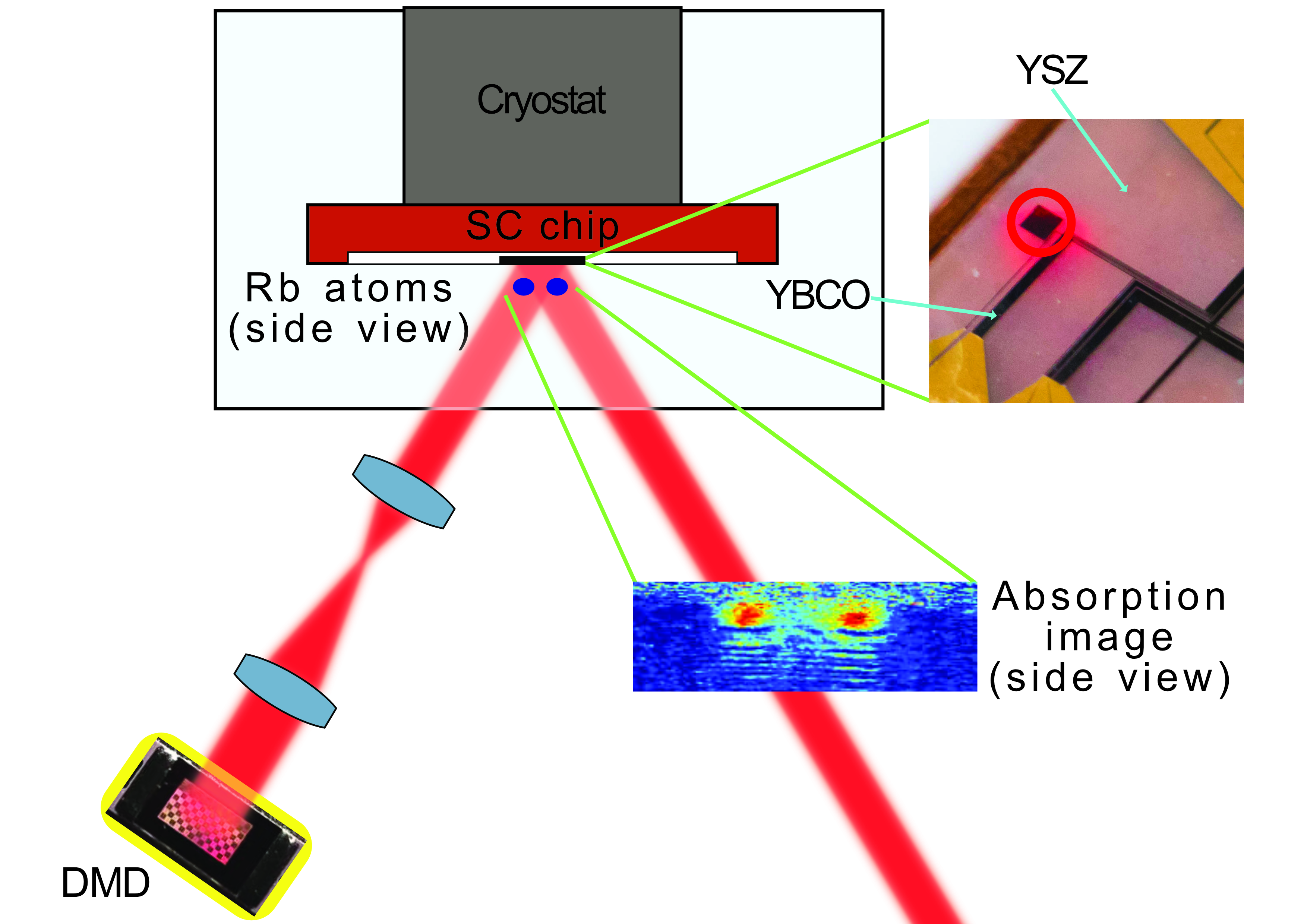}
  \caption{Schematic diagram of the setup. A Gaussian beam of a high power laser is patterned with a DMD and imaged onto a superconducting chip attached to a cryostat inside vacuum. The vortex distribution is probed by absorption imaging \cite{10.1364/OE.19.008471} of the atoms after illumination.}
  \label{fig:laser}
\end{figure}
We prepared an ensemble of ultracold $^{87}$Rb atoms in a standard magneto-optical trap. After molasses cooling and optical pumping, $1\cdot10^{8}$ atoms were magnetically trapped in the $\ket{F'=2,m_{F}=2}$ ground state. The atoms were then magnetically transported vertically by $3.5$ cm into a close proximity of the SC atom chip before being loaded into the traps created by the trapped magnetic field. 

As the SC chip  we employed a $1\times 1$ mm$^2$ square  thin YBCO (Yttrium Barium Copper Oxide, YBa$_{2}$Cu$_{3}$O$_{7-x}$) film  fabricated on a YSZ (Yttria-Stabilized Zirconia) substrate and having the thickness $d=800$ nm.   
For generating different trap geometries we used a $1064$ nm high-power laser and a DMD to locally heat the SC film. Our setup is shown in Fig. \ref{fig:laser}:  the DMD is illuminated with a 2.5 W Gaussian beam; the pulse time and power are controlled via an acousto-optic modulator. The pattern generated by the DMD is imaged onto the surface of the SC film with a magnification of 0.4. 
The vortices were first induced in the SC film by a pulse of a uniform magnetic field. Subsequently, they were patterned with the shaped illumination pulses for $20-50$ ms to obtain the desired trap geometry. 
\begin{figure}[t!]
  \includegraphics[scale=0.37]{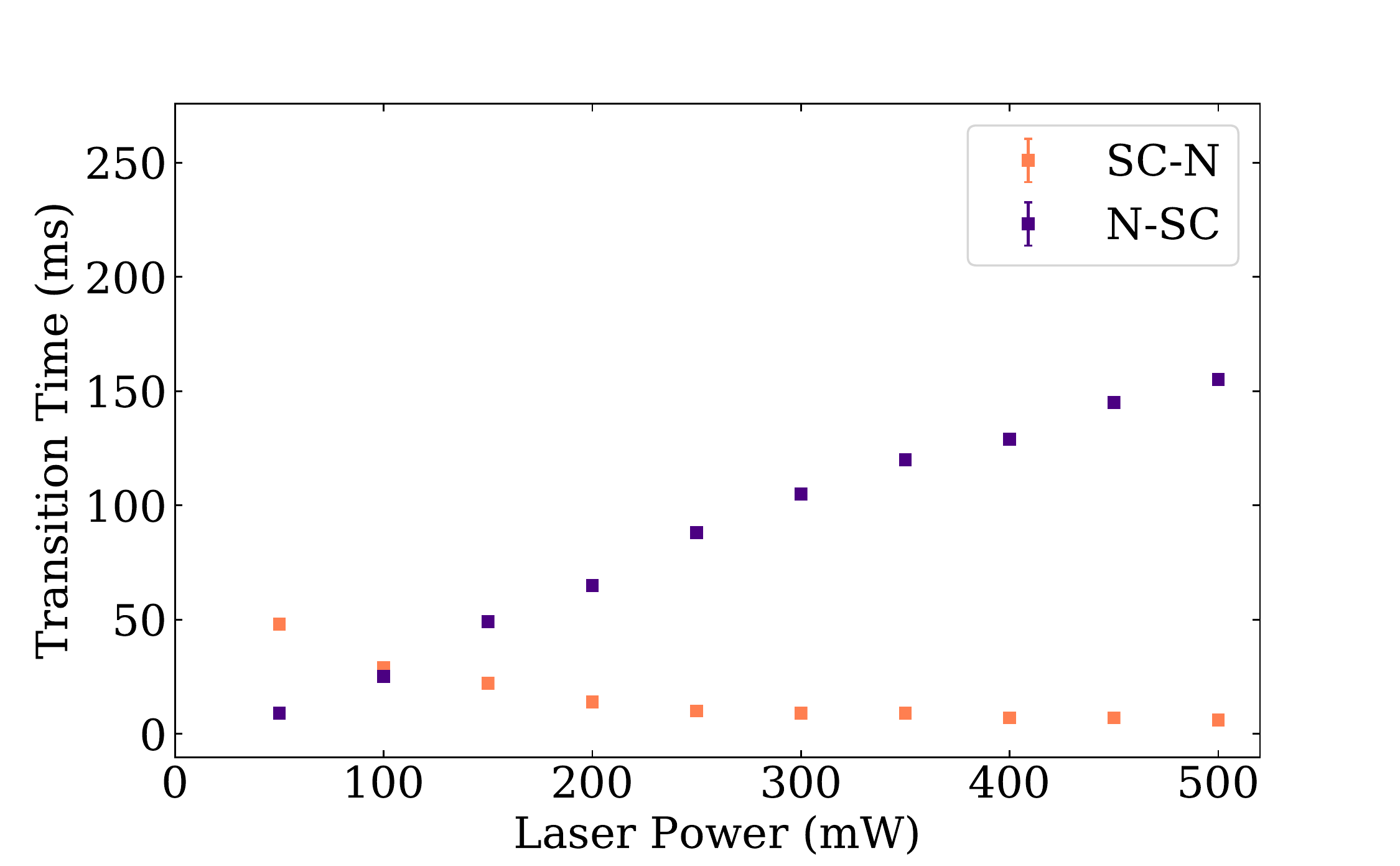}
  \caption{Transition times from SC to N state and from N to SC state when pulsing the laser for $240$ ms.}
  \label{fig:transition}
\end{figure}

The superconductivity in YBCO can be destroyed by optical illumination \cite{PhysRevB.4.2189,92539,doi:10.1063/1.107069}. This process depends on the SC film thickness compared to the optical penetration depth $\delta$, which is about $80-120$ nm for YBCO \cite{234005}. In case of optically thin YBCO films ($d\ll \delta$), the Cooper pairs break on a typical timescale of picoseconds, giving rise to a fast non thermal optical response \cite{doi:10.1063/1.101538,doi:10.1063/1.101539,doi:10.1063/1.106500}. In our case ($d\gg \delta$), the heat is initially deposited in a surface film layer (order of $\delta$), where the Cooper pairs breaking takes place, but then rapidly spreads (in several ps) via non-equilibrium hot electrons deeper into the sample. These electrons thermalize with the lattice (during and after the exposure time) due to electron-phonon interactions and reproduce a typical thermal response. Usually the cool down process 
is much slower because of equilibrium thermal conduction (via phonons) into the substrate. 

To effectively manipulate the vortex distribution we then need to consider the surface temperature rise due to the initial heat absorption and heat diffusion. For laser pulses of milliseconds, the increase in surface temperature can be described by a heat diffusion equation \cite{795571}. For a square pulse, the surface temperature rise $T$ can be estimated following the rate equations in \cite{phdthesis}. For removing all the vortices from a designated area its temperature has to reach the critical value $T_c=87$ K; however, the critical current density drops sharply in the temperature region from $80$ K to $90$ K \cite{doi:10.1002/9783527633357.ch4}. This behavior affects the trapping potential generated for $T<T_c$.  
 
The transition times for the second order phase transition from the SC state to the normal (N) state and reverse are shown in Fig. \ref{fig:transition}. It is evident that the timescales are asymmetric.  Increasing the laser power leads to a decrease of transition time from the SC to N state. The heat transfer from the SC film to the substrate determines the transition time from the N state to the SC state. At laser power greater than $200$ mW there is no change in the SC-N transition  time at timescales relevant to the experiment. However, the N-SC transition time increases  approximately linearly with the laser power.

Designing an optical pattern for a SC film, one has to take into account the heat diffusion. A too long exposure time can smear out the heat pattern. 
\begin{figure}[t!]
\includegraphics[scale=0.37]{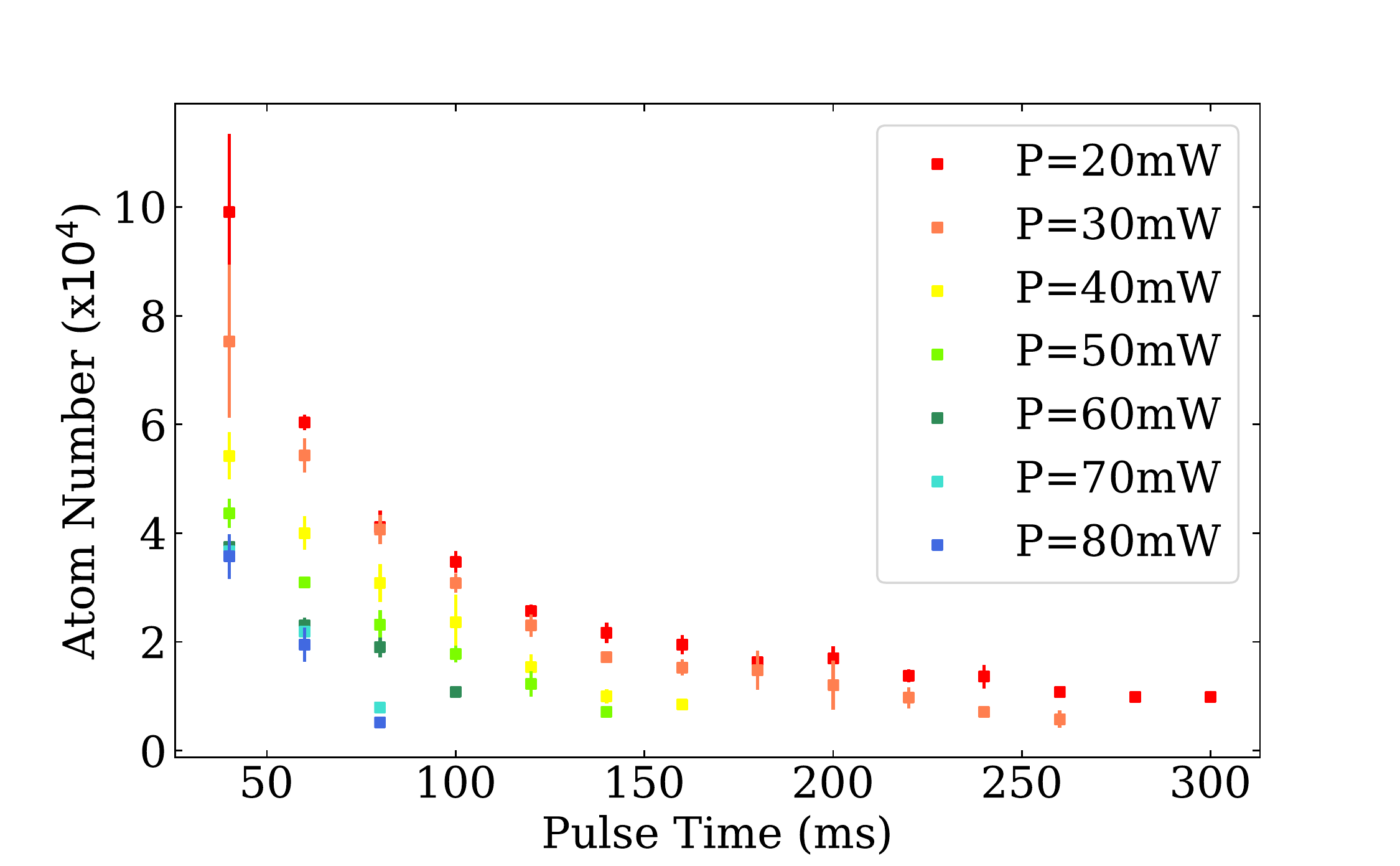}
\includegraphics[scale=0.37]{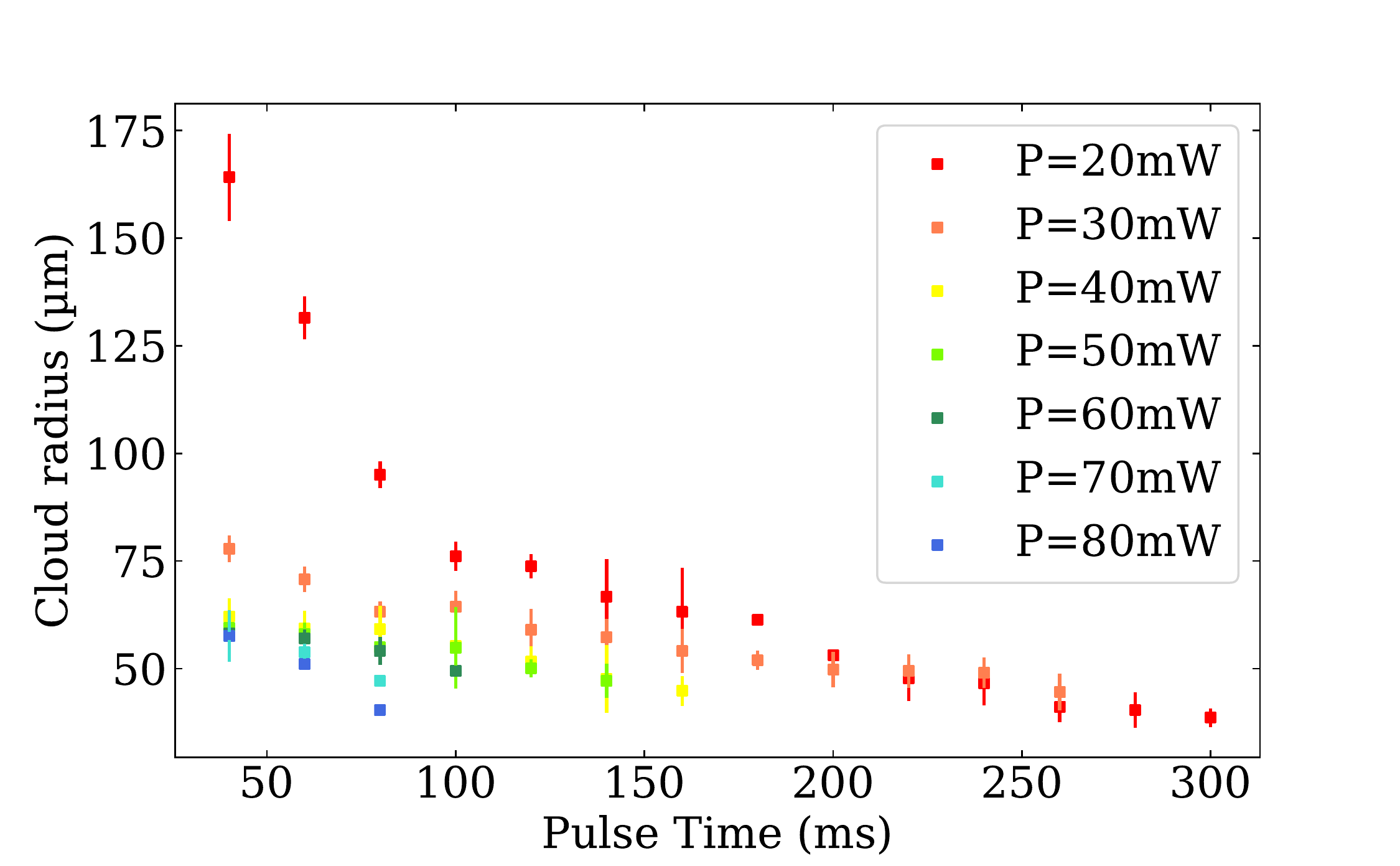}
 \caption{Atoms number and cloud dimension as a function of the half-square illumination pulse time for different beam powers.}
\label{fig:atom} 
\end{figure}
To investigate the heat diffusion, we trapped the atoms by the vortices induced  in the square film and then  illuminated half of the square surface with different laser power. Then the remaining atom number was detected. At low power only the vortices in the illuminated part were affected.  As the power grows, heat  diffusion to the rest  of the film  changes the vortex distribution there too. Both the trapped atom number and the  cloud size decrease as the time of illumination increases (Fig. \ref{fig:atom}). Our simulations and experimental  data on the atom number, the cloud size and location indicate  that the trap depth changes   not only due to the shrinkage of the SC part of the film but also due to decrease of the critical current density. For the pulse power $20$ mW and a $10\%$ loss from reflection \cite{2005PhDT.......225C},   the heat diffusion equation predicts about $5$ K  rise in temperature. This temperature is still below $T_c$, but the critical current changes significantly in this temperature range, resulting in a different trapping potential and in a smaller cloud for longer times. 

\begin{figure}[t!]
\includegraphics[scale=1.5]{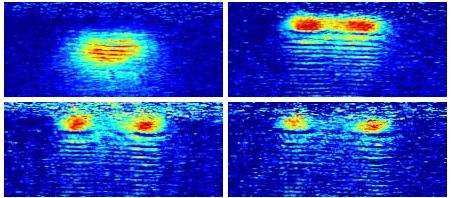}
\caption{Absorption images of the atomic cloud splitting taken after $0$ (top-left), $20$ (top-right), $30$ (bottom-left) and $40$ (bottom-right) ms of illumination time respectively. The interference patterns visible below the atoms are common in imaging systems where the imaging light is reflected from the chip surface before being collected by the camera \cite{Smith_Absorption_Imaging}.}
\label{fig:splitting}
\end{figure}

In one of our experiments we transferred the cold atoms into a magnetic trap created below the magnetized square SC film and then illuminated  a  $1\times 0.1$ mm$^2$ strip centrally placed on the film surface.  This changed the distribution of vortices and the original single atom cloud splitted into two separate clouds (see Fig. \ref{fig:splitting}) .  \\
To simulate the SC film magnetization induced by the applied field variations and manipulated by laser heating,  we used the Fast Fourier Transform (FFT) based numerical method proposed in \cite{0953-2048-25-10-104001} and  modified in \cite{0953-2048-31-5-055018}. The SC film was characterized by the power current-voltage law,  $\bm{e}=e_c(|\bm{j}|/j_c)^{n-1}\bm{j}/j_c$, where $\bm{e}$ is the parallel to film component of the electric field, $\bm{j}$ is the sheet current density, $e_c=10^{-4}$ V$\cdot$m$^{-1}$. The critical sheet current density of the film was estimated as $j_c=160$ A$\cdot$cm$^{-1}$ and $n=50$. The  high value of the power $n$ made the film magnetization model close to the rate-independent critical-state Bean model \cite{RevModPhys.36.31,10.1103/PhysRevB.49.9802}. 

Although the real sequence of applied magnetic field pulses in our experiment was more complicated because of the necessity to bring atoms to the vicinity of the film, for a qualitatively realistic trap simulation it was sufficient to account only for the perpendicular to film applied field which  first changed from zero to 239 G (which is below the full penetration field \cite{PhysRevA.81.053624}), then decreased to -23 G and remained such during the laser heating. We also replaced the realistic heat diffusion model by a simplified one in which heating causes a sheet critical current density drop in a chosen region. Thus, modeling the central strip heating, we assumed that on the last stage of our simulation the sheet critical current density gradually dropped to $0.05j_c$ uniformly in the strip $|x|<0.1$. \\
\begin{figure}[t!]
\includegraphics[width=8.6cm]{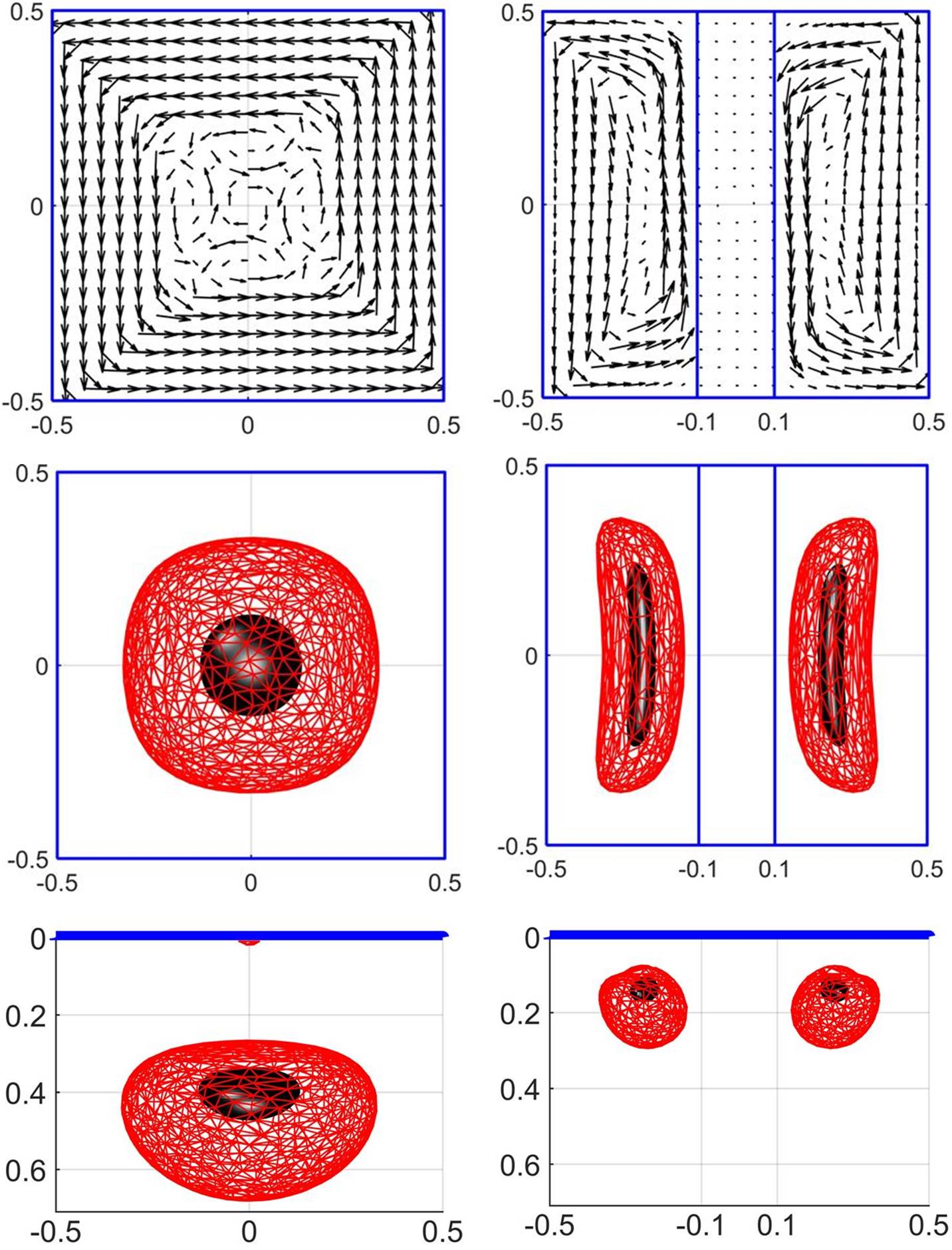}
 \caption{Simulation results: before (left) and after (right) heating the strip $|x|<0.1$. Top: the sheet current density distributions in the SC square. Magnetic traps are represented by the iso-surfaces $|b|=6$ G (black) and $15$ G (red) viewed from below (center) and from the side (bottom).}
\label{fig:current} 
\end{figure}
We noted that the computation domain for the FFT-based method should be larger than the film area; in our simulations the film was centrally placed in the domain $-1.5\leq x,y\leq 1.5$ mm and, for the  uniform  256$\times$256 grid in this domain, the computation took from a few minutes to half an hour on a standard personal computer. 
Computed film current densities before and after strip heating are shown in Fig. \ref{fig:current} (top).
Under the adiabaticity assumption the atom cloud shape can be approximated by the closed iso-surface of the magnetic field magnitude, $|\bm{b}|$,  chosen in accordance with the atom cloud temperature. The magnetic field in a vicinity of the film with a known sheet current density distribution has been computed numerically using the Biot-Savart law and the FFT algorithm. 
By the Biot-Savart law magnetic induction is
$$\bm{b}(\bm{r},t)=\bm{b}_a(t)+\frac{\mu_0}{4\pi}\nabla\times\int_{\Omega}\frac{\bm{j}(\bm{r}',t)}{|\bm{r}-\bm{r}'|}\,d\bm{r'},$$ where $\bm{b}_a(t)$ is the uniform applied field, $\mu_0$ the magnetic permeability of vacuum, $\bm{r}=(x,y,z)$ and $\Omega$ is the two-dimensional film domain in the plane $z=0$. Extending the sheet current density by zero outside the film we can present this field using two-dimensional convolutions:
\begin{eqnarray*}
b_x=\frac{\mu_0z}{4\pi}j_y*\left(\frac{1}{|\bm{r}|^3}\right),\ \ \ \ \ \ 
b_y=-\frac{\mu_0z}{4\pi}j_x*\left(\frac{1}{|\bm{r}|^3}\right),\\
b_z=b_{a,z}+\frac{\mu_0}{4\pi}\left[ 
j_y*\left(\frac{\partial}{\partial x}\frac{1}{|\bm{r}|}\right)-j_x*\left(\frac{\partial}{\partial y}\frac{1}{|\bm{r}|}\right)
\right],
\end{eqnarray*}
where we assumed $b_{a,x}=b_{a,y}=0$. Applying the Fourier transform $F[f]=\int_{R^2}f(x,y){\rm exp}(-{\rm i}\{xk_x+yk_y\})\,dx\,dy$ and taking into account that 
$$F\left[\frac{1}{|\bm{r}|}\right]=\frac{2\pi}{|\bm{k}|}e^{-|z\bm{k}|},\ \ \ 
F\left[\frac{1}{|\bm{r}|^3}\right]=\frac{2\pi}{|z|}e^{-|z\bm{k}|},$$
where $\bm{k}=(k_x,k_y),$ we obtain
\begin{eqnarray*}
b_x=\frac{\mu_0z}{2|z|}F^{-1}\left[e^{-|z\bm{k}|}F[j_y]\right],\\ 
b_y=-\frac{\mu_0z}{2|z|}F^{-1}\left[e^{-|z\bm{k}|}F[j_x]\right],\\
b_z=b_{a,z}+\frac{\mu_0}{2}F^{-1}\left[\frac{{\rm i}e^{-|z\bm{k}|}}{|\bm{k}|}\left(k_xF[j_y]-k_yF[j_x]\right)\right].
\end{eqnarray*}
By the FFT-based method \cite{0953-2048-25-10-104001,0953-2048-31-5-055018} the sheet current density $\bm{j}$ is found in the nodes of a uniform grid. 
Replacing the continuous Fourier transform and its inverse by their discrete analogues on the same grid and using the FFT algoritm, one can efficiently calculate the magnetic field in a vicinity of the film.

The simulated magnetic trap and its split due to optical heating of a strip are shown in Fig. \ref{fig:current} (center and bottom). Figs. \ref{fig:splitting} and \ref{fig:current}  show that the experimental and simulated results agree sufficiently well.  
When no light is applied, there is only one cloud below the center of the film; the cloud size and position are close in the experiment and simulation.

For strip illumination duration of $t=20$ ms there appears a normal region in which the temperature reaches $T_c$ and all vortices are gone. As the exposure time increases, the normal region grows, pushing the centers of the two traps to the edges of the square and decreasing the trap depths; the critical current density decreases and the normal region expands. For $t=100$ ms, no atoms remain trapped. From simulations (magnetic field and current distribution), after $100$ ms an area of $1\times 0.6$ mm$^2$ of the  film reaches the critical temperature  $87$ K leaving two SC regions, each  $~200$ {\textmu}m wide. The temperature in these two regions is below $T_c$ (from $85$ K to $87$ K) but the critical current density is significantly reduced, implying a smaller trapping volume and a lower trap depth.  
 
The  patterned film heating is limited by the optical resolution and heat diffusion. To minimize the latter, the laser intensity and the exposure time have to be chosen carefully. Our current experimental setup limits the pattern resolution to $10$ {\textmu}m.

\begin{figure}[t!]
    \includegraphics[width=8.6cm]{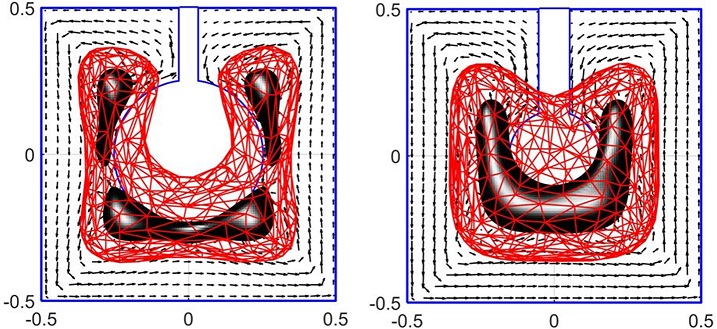}
    \caption{Simulation results. We could obtain quite different trap shapes:  a crescent-like trap or a ring with defects. Here, prior to heating, the external field was first increased to 200 G, then decreased to -10 G; the traps are indicated by the  iso-surfaces $|\bm{b}|=3$ G (black) and 6 G (red).}
    \label{fig:sim}
\end{figure}

Within the experimentally accessible parameter range we have explored various trapping potentials. Examples include trap arrays, rings and rings with defects.  
\begin{figure}[t!]
\includegraphics[scale=0.467]{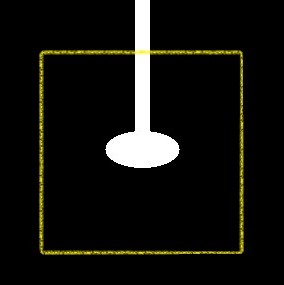}
\includegraphics[scale=2.345]{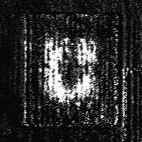}
\caption{DMD image on the left, where the real dimension of the SC film is highlighted in yellow and absorption image of the atomic cloud on the right.}
\label{fig:ring}
\end{figure}
For example, for a light pattern as in Fig. \ref{fig:ring}, left, realized with the DMD projecting the light onto the chip  under a small angle, the atom cloud has the broken ring shape, Fig. \ref{fig:ring}, right.  This situation was simulated numerically for two heated zones, see   Fig. \ref{fig:sim}. Qualitatively, the experimental results were reproduced.  The considered trap configuration can be employed in a superconducting quantum interference device using ultracold atoms (atom SQUID) \cite{1367-2630-19-2-020201,PMID:24522597}. \\
In conclusion, our experimental results and simulations showed that magnetic traps on SC chips can be optically manipulated to tailor the structure and shape of ultracold atom clouds. Various cloud configurations have been realized  experimentally and simulated numerically.

More complex structures can be achieved by increasing the heating pattern resolution. This method can be used to create magnetic trap lattices for ultracold atoms \cite{10.1103/PhysRevLett.111.145304} in quantum computing applications. Optically manipulated SC chips open new possibilities for ultracold atoms trapping and design of compact on-chip devices for investigation of quantum processes and applications in atomtronics.


\end{document}